\documentclass[10pt]{iopart}
\usepackage{graphicx}
\usepackage[normalem]{ulem}

\usepackage{color}

\begin{document}

\title[A Artaud \textit{et al}]{Depressions by stacking faults in nanorippled graphene on metals}

\author{Alexandre Artaud$^{1,2}$, Estelle Mazaleyrat$^{1,2}$, Georg Daniel F\"{o}rster$^3$, Charl\`{e}ne Tonnoir$^1$, Bruno Gilles$^4$, Philippe David$^2$, Val\'{e}rie Guisset$^2$, Laurence Magaud$^2$, Florent Calvo$^5$, Claude Chapelier$^1$ and Johann Coraux$^2$}
\address{$^1$ Univ. Grenoble Alpes, CEA, IRIG, PHELIQS, 38000 Grenoble, France}
\address{$^2$ Univ. Grenoble Alpes, CNRS, Grenoble INP, Institut NEEL, 38000 Grenoble, France}
\address{$^3$ Laboratoire d'\'{E}tude des Microstructures (LEM), CNRS, ONERA, Universit\'{e} Paris-Saclay, 29 Avenue de la Division Leclerc, F-92322, Ch\^{a}tillon, France}
\address{$^4$ Univ. Grenoble Alpes, CNRS, Grenoble INP, SIMAP, 38000 Grenoble, France}
\address{$^5$ Univ. Grenoble Alpes, CNRS, LIPhys, 38000 Grenoble, France}
\ead{johann.coraux@neel.cnrs.fr}

\begin{abstract}

A broad variety of defects has been observed in two-dimensional materials. Many of these defects can be created by top-down methods such as electron irradiation or chemical etching, while a few of them are created along bottom-up processes, in particular during the growth of the material, in which case avoiding their formation can be challenging. This occurs \textit{e.g.} with dislocations, Stone-Wales defects, or atomic vacancies in graphene. Here we address a defect that has been observed repeatedly since 2007 in epitaxial graphene on metal surfaces like Ru(0001) and Re(0001), but whose nature has remained elusive thus far. This defect has the appearance of a vacant hill in the periodically nanorippled topography of graphene, which comes together with a moir\'{e} pattern. Based on atomistic simulations and scanning tunneling microscopy/spectroscopy measurements, we argue that such defects are topological in nature and that their core is a stacking fault patch, either in graphene, surrounded by loops of non-hexagonal carbon rings, or in the underlying metal. We discuss the possible origin of these defects in relation with recent reports of metastable polycyclic carbon molecules forming upon graphene growth. Like other defects, the vacant hills may be considered as deleterious in the perspective of producing high quality graphene. However, provided they can be organized in graphene, they might allow novel optical, spin, or electronic properties to be engineered.
\end{abstract}

%
\vspace{2pc}

\submitto{\TDM}
%
%

\ioptwocol

\section*{Introduction}

Epitaxial ultrathin films often form superstructures due to the lattice mismatch or disorientation with their substrate. A peculiar kind of ultrathin films, two-dimensional (2D) materials like graphene \cite{grant1970,vanbommel1975,zi1987,land1992}, boron nitride \cite{corso2004,laskowski2007}, or transition metal dichalcogenides \cite{kim2011,sorensen2014} are no exception. Some of these superstructures are called moir\'{e} patterns, in an analogy with optics. In the case of 2D materials, the strong internal bonds together with low bending rigidity are responsible for a natural rippling, whose in-plane period is the moir\'{e} pattern lattice parameter, \textit{i.e.} a few nanometers given the typical lattice mismatch with the substrate. Borrowing now the language of geology, the moir\'{e} patterns of 2D materials are associated with a hill-and-valley topography.

The relative height of the hills and valleys depends on several factors. Especially, a stronger interaction between the 2D material and the substrate imposes lower valleys, as extensively discussed in the case of graphene on metals \cite{wang2008,busse2011}. The distinctive bending and binding at hills and valleys have numerous consequences on the properties of graphene (and, presumably, of other epitaxial 2D materials, too). For instance, the electronic band structure of graphene develops replicas of Dirac cones and mini-bandgaps \cite{pletikosic2009}, its work function is larger at the location of the hills \cite{altenburg2014}, its chemical reactivity is different at hills and valleys \cite{altenburg2010,boneschanscher2012,voloshina2013}. Disorder in the hill-and-valley topography may thus have important consequences on these properties.

\begin{figure*}[!hbt]
\centering
\includegraphics[width=83.7mm]{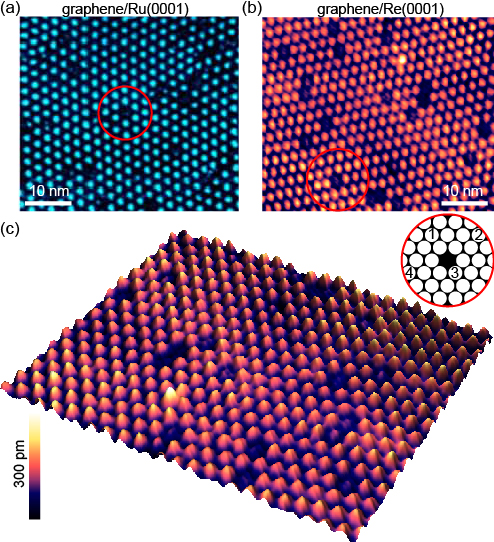}
\caption{Scanning tunneling topographs of graphene prepared on (a) Ru(0001) and (b) Re(0001), revealing a significant density of defects (circled in red) in the moir\'{e} lattice. (c) Three-dimensional view of (b), highlighting the hill-and-valley nanorippled surface associated with the moir\'{e} lattice, and the defects appearing as vacant hills. A closed-path "1-2-3-4" around one vacant hill is schematized in the inset. \textit{a} is reproduced with permission from ref.~\cite{lu2012}.}
\label{fig1}
\end{figure*}

One specific kind of disorder is often encountered in graphene on metals, namely a depression in the pattern of hills spawned by the moir\'{e} (figure~\ref{fig1}). This defect, which we call a "vacant hill", is seen in most scanning tunneling micrographs, whatever the measurement conditions \cite{marchini2007,yi2007,deparga2008,pan2009,sutter2009,donner2009,feng2011,guenther2012,natterer2012,lu2012}, of graphene grown on one of the prominent metal surfaces used to produce large-area high quality graphene, Ru(0001) \cite{sutter2008,sutter2009}. The same defect is observed in graphene grown on Re(0001) \cite{tonnoir2013,qi2017}, and possibly as well on Rh(111) \cite{wang2011,sicot2012}, two similar systems in terms of graphene-metal interaction strength \cite{miniussi2011}. However, to our best knowledge, it has not been seen on other metal surfaces.

Much like with any other defect, reducing the density of vacant hills is desirable for many applications. This is presumably needed to achieve pure band-like electronic and phononic properties, or to increase the degree of ordering in 2D lattices of nanoparticles self-organized on the graphene/Ru(0001) or graphene/Rh(111) moir\'{e}s \cite{donner2009,pan2009b,sicot2010,liao2011,engstfeld2012,wu2013,teng2014,xu2014,foerster2015}. Additionally, vacant hills were suggested to be entry points for the intercalation of oxygen between graphene and Ru(0001) \cite{lu2012}. This is obviously a shortcoming in the prospect of fully impermeable membranes or protective coatings. Yet such pathways have already been exploited to decouple graphene from its substrate and to create highly stretched graphene nanobubbles with strong, Landau-level-like intrinsic electronic resonances \cite{lu2012}.

The question that we focus on here has remained open since the first scanning tunneling microscopy (STM) images of graphene/Ru(0001) were published in 2007 \cite{marchini2007,yi2007}, and a few years later in graphene/Rh(111) \cite{wang2011} and graphene/Re(0001) \cite{tonnoir2013}: What is the nature of the vacant hill? The answer to this question is missing because the available observations are unable to reveal the atomic structure. Our strategy is to consider different classes of atomic structures involving deviations to the perfect atomic lattices of the metal surface or graphene. Such a survey has been so far out of reach of density functional theory (DFT) calculations in reason of the prohibitive computing time required to address already a few of the candidate structures. Instead, we use molecular dynamics simulations based on a parametrized bond-order potential (BOP) to explore a variety of structures, and DFT calculations too in the case of simple structures. The confrontation of the simulation results to high resolution STM and scanning tunneling spectroscopy (STS) data leads to the conclusion that at the center of the vacant hills, the stacking of carbon atoms on the substrate differs from that found at a regular hill. The vacant hill may then be seen as a stacking fault where bonding to the substrate is stronger. Possible structures feature a stacking fault either in the metal surface or in the graphene lattice. In the latter case, loops of defects stitch the stacking fault to the surrounding graphene, and these loops comprise non-hexagonal carbon rings. In this sense, the vacant hills are topological defects and their electronic properties naturally differ from those of regular hills.

\section*{Methods}

Graphene growth was prepared on thin Re(0001) films on sapphire (figure~\ref{fig1}(b,c),\ref{fig5}) and on a Re(0001) single crystal (figure~\ref{fig2}).

Rhenium films with 30~nm-thickness were grown in a dedicated ultrahigh vacuum system using an electron-beam evaporator at a rate of 15~\AA/min at 1200~K. Pieces of $\alpha$-Al$_2$O$_3$(0001) wafers, annealed under ultrahigh vacuum at 650~K for 4~h beforehand, were used as growth substrates. Then the thin film samples were transported in atmospheric conditions into another ultrahigh vacuum system, where they were first annealed to 1070~K for 30~min. To grow graphene, the sample surface was then exposed to 5$\cdot10^{-7}$~mbar of ethylene at 1260~K for 20~min, and subsequently cooled down to 970~K within 50 min. This procedure yields single-layer graphene covering the whole sample surface \cite{tonnoir2013}.

The (0001) surface of a Re single crystal (Surface Preparation Laboratory) was cleaned in ultrahigh vacuum (base pressure $\sim$10$^{-10}$ mbar) by repeated cycles of 2~keV Ar$^+$ ion bombardment at 1020~K and flash annealing to $\sim$1570~K. Graphene was prepared by first saturating the Re(0001) surface with ethylene at room temperature and then two subsequent flash annealings to 1020~K with a $5\cdot10^{-7}$~mbar partial pressure of ethylene \cite{miniussi2014}.

STM measurements were performed in two separate systems. In the same ultrahigh vacuum system where graphene was grown, STM was performed at room temperature using a commercial Omicron UHV-STM 1, with W chemically-etched tips. The samples were also transported out of ultrahigh vacuum, in a home-made STM/STS setup implemented inside a $^{4}$He refrigerator cooled down to 4~K. The differential conductance spectra were acquired using a standard lock-in technique with a 5~mV modulation of the tip-sample bias voltage ($V_\mathrm{t}$) at 480~Hz.

Ab initio calculations were performed using  VASP \cite{kresse1993}. The PAW approach \cite{kresse1999}, PBE functionnal \cite{perdew1996}, and Grimme corrections to van der Waals interactions \cite{grimme2006} have been used. The  system is described by a slab that contains five Re layers, one graphene layer, and a 10~\AA-thick empty space on top. Atoms in the third Re plane were fixed while all the other atoms were let free to relax. The in-plane size of the supercell used for the calculation was set by assuming a coincidence of (8$\times$8) graphene unit cells on top of (7$\times$7) Re(0001) unit cells, corresponding to the experimental observations. A single $k$ point, the supercell K point, was used to get a precise description of the graphene low energy states. After convergence, residual forces were less than 0.025~eV/\AA.

Classical simulations were performed with the bond-order potential \cite{foerster2015}, combining the original BOP of Brenner for carbon \cite{brenner1990} and the embedded-atom model of Li \textit{et al.} \cite{li2004} for ruthenium. In addition, the carbon-metal parameters were adjusted based on DFT data on graphene/Ru(0001) available in the literature \cite{wang2010}. Charge transfers, such as those possibly occurring at the location of defects in graphene or the metal, are implicitly accounted for. Two versions of the BOP were originally designed, with and without dispersion corrections. Geometric properties predicted by DFT being in slightly better agreement with the latter, this version was chosen in the present work.

Lateral periodic boundary conditions are applied with a box dimension chosen such that the lateral graphene lattice constant is imposed at 2.497~\AA\;\cite{wang2010}. The simulations were then conducted on the 12 on 11 moir\'e commensurability, where 12 graphene unit cells match 11 unit cells of the Ru(0001) surface lattice, thus setting the $hcp$ lattice constant $a$ of the metal to be 2.724~\AA, letting the other lateral constant as free to accomodate better upon geometry optimization. Only three layers of ruthenium in ABA stacking were explicitly considered in all calculations.

The systems considered were vacant hills with different structure ($1\times 1$ supercell) possibly surrounded by regular hills (up to $4\times 4$ supercells in total), or a maximum number of 10\,416 atoms. After a short molecular dynamics trajectory at 300~K, the geometry of the entire system was locally minimized, and its structural details scrutinized.

\section*{Nanorippling and relationship with atomic stacking}

Before addressing the vacant hill defect in itself, we briefly remind elementary geometrical features of the moir\'{e} pattern between graphene and a metal surface (detailed descriptions may be found in focused reports \cite{hermann2012,zeller2014,artaud2016,zeller2017}). Disregarding first the nanorippling of graphene, the relative in-plane stacking of C atoms on the metal lattice varies from site to site in the moir\'{e} pattern due to the lattice mismatch and disorientation. Three high-symmetry stacking types are usually considered, namely $fcc$-$hcp$, where the two C sublattices of graphene each sit on different hollow binding sites (usually called $hcp$ and $fcc$) of the substrate, $top$-$hcp$ and $top$-$fcc$, where one of the C sublattices sits directly atop a metal atom and the other on a hollow site ($hcp$ or $fcc$). For these three kinds of stacking the interaction between graphene and the substrate is different as well \cite{wang2008,miniussi2011,busse2011}, thus the graphene-metal distance is different: this is what drives the nanorippling of graphene.

\begin{figure*}[!hbt]
\centering
\includegraphics[width=84.8mm]{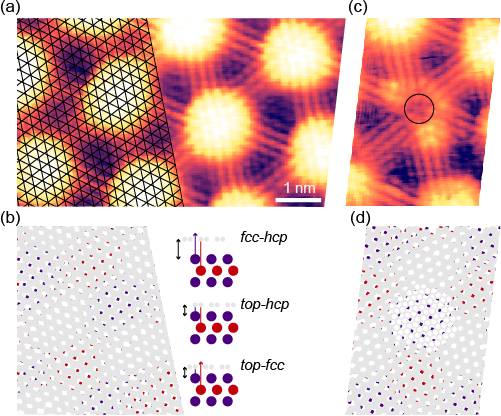}
\caption{(a,c) Atomic-resolution scanning tunneling topographs of graphene on Re(0001), revealing the moir\'{e} pattern and a vacant hill defect (c). Part of (a) is overlaid with a grid whose vertices match the positions of C atoms of one of the two graphene sublattices. (b,d) Positions of the C atoms (grey balls) extracted from (a,c). In (d), the white balls with grey contour are the positions of C atoms extrapolated from the surrounding graphene lattice, \textit{i.e.} in the absence of the vacant hill defect. The atomic positions in the two topmost Re layers have been refined to match the moir\'{e} pattern in (a,c) (hills appear in blue). The C/Re stackings for the three characteristic sites of the moir\'{e}, the two valleys ($top$-$hcp$, $top$-$fcc$) and the hill ($fcc$-$hcp$), are sketched with three side views.}
\label{fig2}
\end{figure*}

These considerations help to rationalize the atomic resolution STM observations (figure~\ref{fig2}(a)) of the moir\'{e} pattern for graphene on Re(0001). On top of the STM image in figure~\ref{fig2}(a), a grid whose vertices mark the positions of C atoms belonging to one of the two sublattices of graphene was overlaid \cite{coraux2008,artaud2016,noteongrid}. To produce a geometrical construction of the moir\'{e} matching figure~\ref{fig2}(a), the two C sublattices were displayed and the periodicity and rotation of the two topmost planes of metal atoms were adjusted, both planes being simply in-plane shifted according to the hexagonal compact stacking in Re(0001). The result is shown in figure~\ref{fig2}(b). At the location of the $fcc$-$hcp$ stacking, \textit{i.e.} the location of a moir\'{e} hill, we only see through the graphene lattice the Re atoms of the topmost layer. At the location of the $top$-$hcp$ and $top$-$fcc$ stackings (which are two inequivalent moir\'{e} valleys), respectively no Re atom and the atoms from the second Re plane are seen through. Note that the present discussion can be directly transposed to the case of graphene on Ru(0001).

We now turn to atomic resolution imaging at the location of a vacant hill. A typical image is shown in figure~\ref{fig2}(c). It resembles the one published in ref.~\cite{qi2017}. Here we attempt to determine the positions of C atoms, just like what was achieved for graphene on Re(0001) in the absence of defects. Away from the defect site this task is rather straightforward. In contrast, at the defect location we are faced with the usual shortcomings of STM imaging: the details that we observe are both of topographic and electronic nature, and without \textit{a priori} knowledge on the structure of the defect, determining the atomic positions is ambiguous. We therefore temporarily refrain to devise on the detailed nature of the defect. This is why in figure~\ref{fig2}(d), the carbon atoms are represented as open symbols at the location of the defect: these symbols correspond to an extrapolation of the perfect graphene lattice. We note, once more, that if graphene were perfect and the Re unperturbed there, then the C/Re stacking would be of $fcc$-$hcp$ type (Re atoms of the topmost substrate plane seen through).

\section*{Nanometer-scale stacking faults}

One important observation can be made to get further insight into the nature of the vacant hill. Within the circle drawn in figure~\ref{fig2}(c), we discern three to four hollow spots. These spots may correspond to the centers of carbon rings or to one of the two carbon sublattices, which can appear as hollow sites depending on the local graphene/metal stacking \cite{wang2008,artaud2016}. Within the resolution of our image, the spacing between these spots is compatible with the graphene lattice parameter, and the spots seem to align along directions parallel to zigzag C rows in the surrounding graphene. Whether these spots lay  or not on the vertices of the network of highest-symmetry lines of the surrounding graphene cannot be determined given the resolution of our STM image. For these few defect-related spots the triangulation procedure that we used to extract the position of C atoms of the defect-free graphene lattice \cite{noteongrid} becomes unreliable. Other procedures based on Fourier filtering \cite{hytch1998,liu2005,lawler2010,vlaic2018} may be used instead to extract the order parameter field from the STM image. In the present case the order parameter is a two-fold scalar (or a vector) measuring the displacement of the atoms with respect to those of a perfect (strain-free) graphene lattice. We attempted the corresponding so-called geometrical phase analysis \cite{hytch1998,rouviere2005}. To reach the high spatial resolution required here (the core of the vacant hill is hardly 1~nm-large), a rather large region of interest must unfortunately be chosen in the Fourier space, which makes the analysis extremely sensitive to (otherwise negligible) experimental artefacts and prevents the extraction of any meaningful information. 

The observed atomic-scale contrast at the center of the vacant hill suggests that at the center of the defect, the carbon lattice is unrotated with respect to the surrounding graphene, although it is possibly in-plane shifted. This rules out an interpretation of the defect as a rotated nano-grain such as the ones observed in epitaxial graphene grown on silicon carbide \cite{cockayne2011}.

Vacancies in the carbon or metal lattices could be another interpretation. Atomic vacancies in the carbon lattice are for instance known to increase graphene's reactivity, leading to the formation of carbon-metal bonds and a decreased distance between graphene and the metal \cite{ugeda2011,standop2013,blanc2013}. This behavior however depends on the region (\textit{e.g.} $top$-$hcp$, $top$-$fcc$, $fcc$-$hcp$) where the carbon vacancy is located: on a $fcc$-$hcp$ (hill) region, there is no metal atom directly underneath the defect, and both DFT and BOP simulations reveal that the graphene height is not significantly altered there \cite{noteonvacancyC}. Constant height-cuts in the electronic density of states shown in the supplementary information (SI) figure~S1 (\textit{link to SI will be provided by editor}) are strikingly different from the STM images of the vacant hill.

Vacancies in the metal lattice have been invoked in another epitaxial 2D material with vacant hills, namely MoS$_2$ grown on Au(111) \cite{krane2016}. In these vacant hills no atomic defect was found in the 2D material, and their measured low electronic density of states was ascribed to the presence of a nanometer-scale atomically-deep vacancy island in the Au(111) substrate.

\begin{figure*}[!hbt]
\centering
\includegraphics[width=82.5mm]{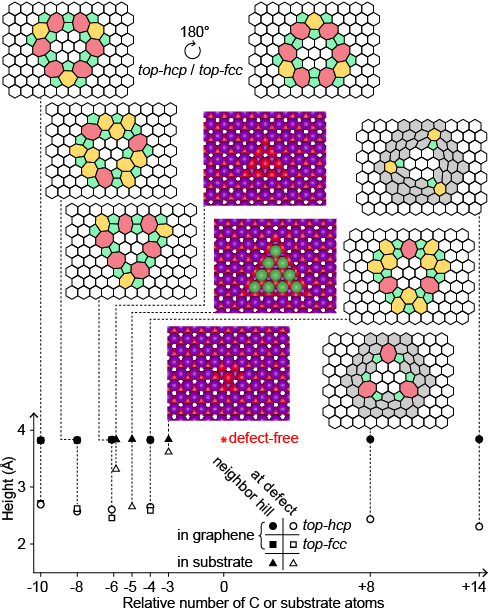}
\caption{Height of graphene with respect to the average height of the uppermost Ru(0001) metal plane (averaged away from the defect location) at the location of a defect (open symbols) and of the neighbor hills (filled symbols), as function of the relative number of C or substrate atoms compared to the defect-free graphene or substrate. Two structures ($top$-$hcp$, $top$-$fcc$) have been considered for four of the defects in graphene (-10, -8, -6, -4 C atoms), rotated by 180$^\circ$ with one another.}
\label{fig4}
\end{figure*}

Using numerical simulations to model such defects (and others in graphene, see below), where the atomic structure of the regular lattice is altered over typically 1~nm, requires large calculation supercells (several thousands of atoms) without which spurious boundary effects would be likely present and difficult to evaluate. This is prohibitive for most DFT approaches. We use instead BOP calculations that allow thousands of atoms to be treated (SI). A BOP potential has in fact been carefully parametrized for the Ru-C system against the results of DFT calculations, and it was found to satisfyingly reproduce the structural parameters of the nanorippled graphene/Ru(0001) moir\'{e} \cite{foerster2015}. Interestingly, (0001)-terminated Ru and Re have strong electronic similarities, with in both cases $d$ bands in the 2--4~eV range below the Fermi level \cite{sutter2009b,papagno2013,estyunin2017}. This is possibly a reason why sp$^2$-hybridized carbon behaves very similarly on Ru(0001) and Re(0001): on both surfaces, similar small size-selective polycyclic clusters tend to form \cite{lu2011,cui2011,artaud2018} and  graphene is nanorippled with almost identical minimum and maximum graphene-metal distances \cite{wang2008,miniussi2011,tonnoir2013,gao2017}. Hence, while an optimized BOP potential is not yet available for Re-C, we expect it to be very similar to the one presently used for Ru-C.

We first rule out the possibility that the vacant hill is a metastable (local minimum) configuration of the graphene layer, where the nanorippling would simply not have occurred. For that purpose we performed a molecular dynamics simulation from an initial structure where a single regular hill was suppressed, by initially locally positioning the C atoms there closer to the metal than they would be in a regular hill. Thermalising the system at 300~K, this local depression spontaneously transforms into a regular hill in less than 1~ps.

\begin{figure*}[!hbt]
\centering
\includegraphics[width=150.0mm]{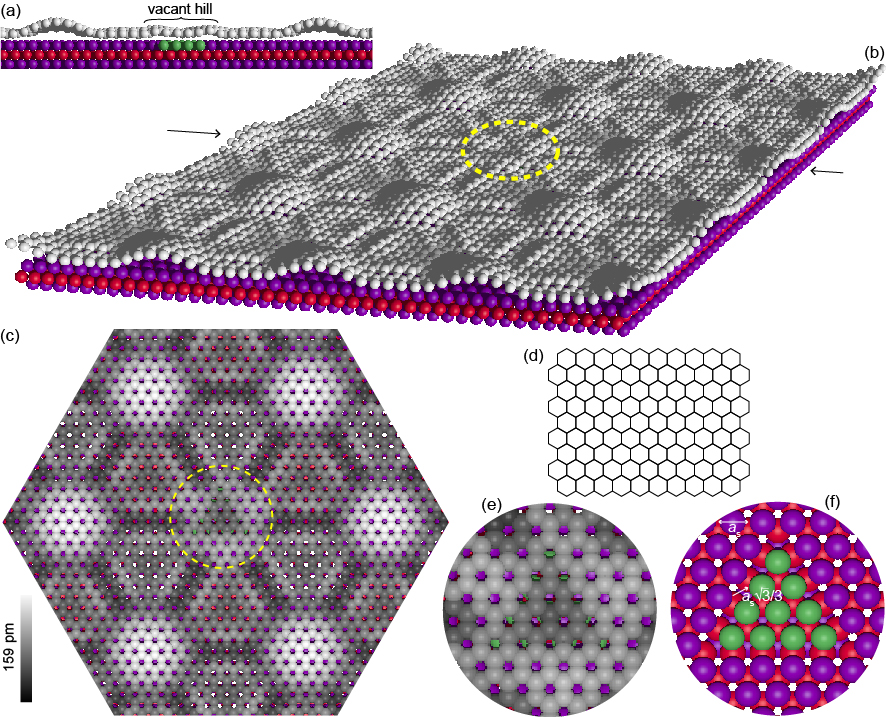}
\caption{Vacant hill model with a stacking fault in the metallic substrate. (a) Side-view ball model of the optimized nanorippled topography of graphene on Ru(0001), with two hills and one vacant hill. The amplitude of the out-of-plane displacement of C atoms has been multiplied by two; (b) Three-dimensional view corresponding to (a). Two arrows mark the cut used in (a); (c) Top-view of the carbon topography, with C atoms shown as balls whose grey shade codes the height of the C atoms above 2.347~\AA\; from the average height of the topmost metal plane; (d) Honeycomb graphene lattice exclusively composed of hexagonal rings at the location of the vacant hill; (e,f) Closer view (yellow dotted circle in (b,c)) on a vacant hill, with and without displaying the top graphene layer, showing that carbon atoms atop the faulted substrate atoms (green balls) bring graphene closer to the substrate.}
\label{fig3bis}
\end{figure*}

We next consider candidate structures where the metal surface is locally modified. We examined two possible vacancy islands in the metal surface, triple- and sextuple-vacancies with triangular shape. Our numerical simulations reveal no vacant hill at the location of the vacancies (see figure~\ref{fig4} and figure~S2). We note that the local graphene/metal stacking corresponds to $top$-$fcc$ at the location of the vacancy, but with a much larger distance than in the valley with the same stacking in defect-free graphene. The distance is close to 3.5~\AA, very close to the height of regular moir\'{e} hills. At the moir\'{e} hills, where graphene is expected to be decoupled from the substrate \cite{wang2008,tonnoir2013}, the local density of states is enhanced (\textit{vide infra}), implying that the apparent height as assessed with STM may be an over-estimation of the true height. In the case of the metal vacancies discussed in the present paragraph (for which the distance is also large), one may expect a protrusion in the STM image, and not a depression ("vacant hill") as we observe.

A straightforward configuration with the same stacking ($top$-$fcc$), but with the metal surface closer to graphene, is a stacking fault nanoisland in the metal uppermost layer. In this defect (points with -5 substrate atoms in figure~\ref{fig4}), the metal atoms (shown as green balls in figure~\ref{fig3bis}) are positioned on $fcc$ hollow sites of the metal, \textit{i.e.} they are laterally shifted by $a_\mathrm{s}/\sqrt{3}$ ($a_\mathrm{s}$ being the substrate's lattice parameter) along a $\left\langle 1\bar{1}00 \right\rangle$ direction. The shorter distance promotes a stronger metal-graphene interaction, and the simulations find a clear local energy minimum for a vacant hill (figure~\ref{fig3bis}), which appears deeper by about 1.2~\AA\; compared to regular hills. There we expect a similar conductance as in the moir\'{e} valleys. This configuration is the first possible structure that we propose for the vacant hill, and we anticipate that stacking fault islands with slightly different shapes or sizes are also potential candidates.

The effects of such defect configurations can now be discussed in relation with STM observations of vacant hills in graphene on Ru(0001). These observations found no defect in the carbon lattice, implying that the vacant hill corresponds to a defect in the Ru(0001) substrate \cite{lu2012}. It was suggested that these defects are metal vacancies, for instance induced by ion beam bombardment during the preparation of the sample surface \cite{lu2012}. We however expect that at the temperature used to grow graphene, the metal surface reconstructs in a vacancy-free, staircase of atomically-flat terraces. Besides, we have just discussed that our simulations do not support a scenario with vacancies, and rather point to stacking faults. Interestingly, stacking fault formation has been reported previously in Ru(0001) upon graphene growth, in this case in the form of a lattice of stacking faults \cite{starodub2009}.

It should be noted that graphene as studied in ref.~\cite{lu2012} has rotational disorder, while we have carefully selected highly-ordered graphene samples in the present work. To prove this, closed Burgers circuits like the "1-2-3-4" circuit in the inset of figure~\ref{fig1}(c) can be drawn around any vacant hill observed in figure~\ref{fig1}(a,b), ruling out the presence of dislocations in the moir\'{e} lattice. Moir\'{e} dislocations would necessarily be observed in the presence of dislocations in the carbon lattice \cite{coraux2008}, and these atomic-scale dislocations accompany rotational disorder in a 2D material like graphene \cite{simonis2002,coraux2008}.

Dislocations are surrounded by mechanical deformations, which alter the regular stacking configuration between the C and metal substrate atoms at the origin of the hill-and-valley topography of the moir\'{e} pattern. It is then conceivable that some of these deformations cause a vacant hill where the otherwise expected $fcc$-$hcp$ stacking is strongly perturbed, even in the absence of a metal stacking fault underneath, and this could be the origin of (at least some of) the vacant hills reported in ref.~\cite{lu2012}. This is the second possible structure that we propose for the vacant hill.

We now discuss a third possible kind of structure for the vacant hill, in which the C atoms' positions are laterally shifted with respect to the regular lattice. In other terms, this nano-grain is a graphene stacking fault, with a $top$-$fcc$ and $top$-$hcp$ stacking on top of the substrate instead of the regular $fcc$-$hcp$ stacking, the substrate being unaltered (to a first approximation) by the presence of the defect. The boundary between such a stacking fault and the surrounding graphene necessarily consists of non-hexagonal rings, forming a loop of defects.

\begin{figure*}[!hbt]
\centering
\includegraphics[width=150.0mm]{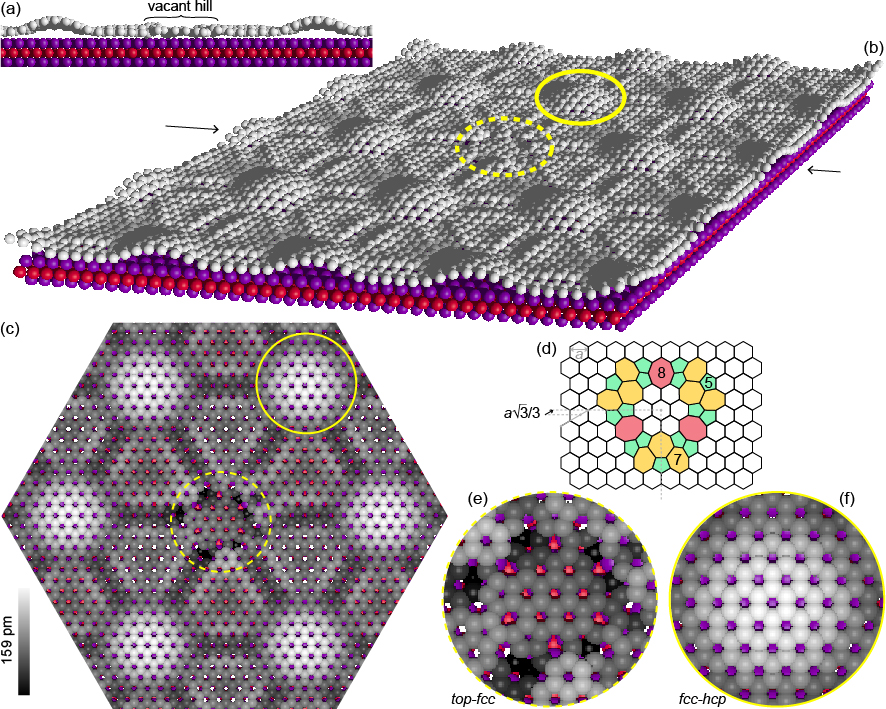}
\caption{Vacant hill model with a stacking fault in graphene itself. (a) Side-view ball model of the optimized nanorippled topography of graphene on Ru(0001), with two hills and one vacant hill. The amplitude of the out-of-plane displacement of C atoms has been multiplied by two; (b) Three-dimensional view corresponding to (a). Two arrows mark the cut used in (a); (c) Top-view of the carbon topography, with C atoms shown as balls whose grey shade codes the height of the C atoms above 2.347~\AA\; from the average height of the topmost metal plane; (d) Details of the C structure of the defect, revealing a lateral shift along the armchair direction of the core of the defect relative to the surrounding defect-free graphene. The grain boundary consists of pentagons, heptagons, and octagons; (e,f) Closer views of \textit{c} (yellow solid and dotted circles in (b,c)), showing the $top$-$fcc$ ($fcc$-$hcp$) stacking of C on Ru at the location of a vacant hill (regular hill) where the second (first) Ru plane is seen through the carbon rings, in red (purple).}
\label{fig3}
\end{figure*}

Here again we use BOP calculations to assess the stability and topography of several such candidate structures. These structures differ by the kind of nano-grain boundaries around the stacking fault at the center of the vacant hill. The stacking fault is a lateral shift of the graphene lattice by $a/\sqrt{3}$ ($a$ being graphene's lattice parameter) along armchair directions, with opposite orientations for the $top$-$fcc$ and $top$-$hcp$ stacking types (see an example in figure~\ref{fig3}(d)).

Strain (shear, stretch, compression) and non-hexagonal rings can accomodate these lateral shifts. As discussed below, strain alone is not sufficient, while non-hexagonal rings alone are. Strain fields extending over several nanometers were found to accomodate stacking faults in a related system, bilayer graphene \cite{alden2013,lin2013,butz2014}. The accommodation of a stacking fault grain with solely strain fields is thus in principle conceivable. However the width across which strain would extend (imposed by the stiffness of C bonds) seems incompatible with the comparatively small size observed for the vacant hill. Besides, a pure-strain stacking fault boundary would not respect the observed symmetry of the defect.

Chains of non-hexagonal carbon rings with two pentagons and one octagon as repeating units \cite{lahiri2010,chen2014}, or tilted Stone-Wales defects (two pentagons and two heptagons) \cite{lehtinen2013}, have already been observed between two-laterally shifted graphene domains. These chains represent discontinuities in the translational order parameter field of the graphene lattice and are therefore one-dimensional topological defects. We then explored the stability of a variety of looped grain boundaries constructed with fragments of these two kinds of topological defects.

To limit the number of possible configurations we focus on those with the highest symmetry and a small core made of six or seven hexagonal rings. The optimized structure of such a defect with a six-hexagon-core and a C density closest (-4 C atoms in figure~\ref{fig4}) to the case of defect-free graphene is represented in figure~\ref{fig3}(a-c) from different viewpoints and is sketched without the substrate in figure~\ref{fig3}(d). The graphene nanograin is surrounded by non-hexagonal rings, here pentagons, heptagons, and octagons. The depression that is observed is a clear energy minimum for the system. This finding confirms that a stacking fault, here in the graphene lattice, naturally gives rise to a vacant hill in the moir\'{e} pattern (clearly apparent when comparing figure~\ref{fig3}(e,f)).

In figure~\ref{fig4} we show alternative candidate structures for the vacant hill defect, with two shapes of nanograins and different kinds of boundaries. Each of these structures can be stacked in a $top$-$hcp$ or $top$-$fcc$ manner on the substrate. These two \textit{alter ego} structures differ by a few 100~meV in binding energy, with a hierarchy that depends on the distance between the defects (see table~S1). It is then hard to judge whether one or the other configuration is preferentially expected.

The BOP simulations predict that the vacant hill is deeper than the neighbor regular hill by typically 1~\AA, and this holds for a rather broad range of C densities lower than that of defect-free graphene. In the case of an excess of C atoms (+8, +14), the graphene topography seems much more complex than in the experimental observations. It consists in strong, localized protrusions and depressions in graphene (above 1~\AA). This is better visualized in figure~S4 than in figure~\ref{fig4}, where the height assessment we use, at the very center of the defect, does not account for the peculiar topography. The (strongly) distorted hexagonal rings of excess-C-structures do not seem relevant ingredients to obtain looped grain boundaries forming vacant hills.

It is tempting to discuss the relative energies of the system with different kinds of defects, to try to identify the most probable structure for the vacant hill. Unfortunately, comparing the total energies of systems comprising a different number of atoms is meaningless. In addition, experimental observations of \textit{e.g.} Stone-Wales defects indicate that defect formation during graphene growth is not forbidden even in case of high, few-electron volt formation energies \cite{banhart2011}. As we will soon see, there are however other arguments, related to the origin of the vacant hill defect, that are more insightful in this respect.

Before we turn to this discussion, we would like to mention the observations of related defects in free-standing single-layer WSe$_2$, that may also be seen as a stacking fault surrounded by loops of non hexagonal rings \cite{lin2015}. These defects are obtained after electron-beam irradiation, that leads to departure of Se from the material. It seems that WSe$_2$ is more tolerant than graphene to the presence of dangling bonds.

\section*{On the possible origin of the vacant hill defects}

Unlike the defects reported in WSe$_2$ \cite{lin2015}, the defects of interest here have formed upon epitaxial growth of graphene. The cores of the vacant hill structures that we discussed above are polycyclic clusters. Alone, these cores remind the size-selective clusters that are specifically observed on Rh(111), Ru(0001), and Re(0001) at early stages of graphene growth \cite{wang2011,cui2011,lu2011,wang2017,artaud2018}. In a recent study we established that these clusters are stacked in a $top$-$fcc$ configuration on the Re(0001) surface, and that they are metastable species whose occurence is kinetically controlled \cite{artaud2018}. It is then tempting to consider the clusters as natural sources of stacking faults in the system, and different viewpoints can be considered whether the stacking fault constituting the vacant hill is in the metal surface or in graphene.

\begin{figure*}[!hbt]
\centering
\includegraphics[width=82.5mm]{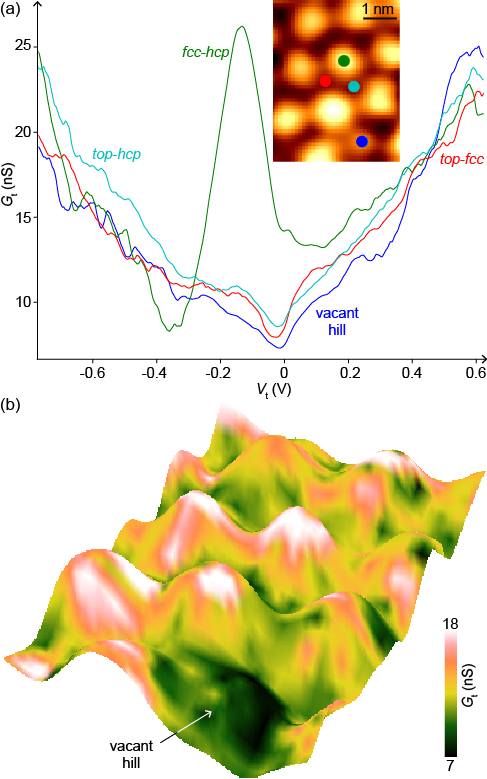}
\caption{(a) Conductance $G_\mathrm{t}(V_\mathrm{t})$ spectra measured at the location of a vacant hill, and of the $fcc$-$hcp$ (hill), $top$-$fcc$ (valley) and $top$-$hcp$ (valley) sites of the moir\'{e}. Inset: STM topograph and the four locations where $G_\mathrm{t}(V_\mathrm{t})$ spectra have been measured; (b) Apparent topography measured with STM in a region comprising a vacant hill, represented with a three-dimensional rendering and overlaid with the corresponding local differential conductance map measured below the Fermi level at $V_\mathrm{t}=-0.1~\mathrm{eV}$. The STM topography is measured on a 2$\times$2 finer grid than the conductance map, which is hence extrapolated.}
\label{fig5}
\end{figure*}

Except where, by chance, a growing graphene patch meets a preformed metastable cluster with locally the same $top$-$fcc$ stacking, the cluster and the graphene patch will be faulted with respect to one another. As we have shown above, this results in the persistence of certain clusters until advanced stages of graphene growth are reached \cite{artaud2018}. The open question at this point is how the cluster is eventually incorporated in the surrounding graphene lattice. This incorporation could occur by the formation of loops of non-hexagonal rings surrounding the graphene nanograin, or by a lateral displacement of the cluster dragging a patch of the topmost metal layer via the strong bonds between the C atoms at the cluster's periphery and the metal surface \cite{artaud2018}, thereby generating a stacking fault in the metal. Note that the latter process is expected to yield metal adatoms from the surface, which could for instance attach to the closest metal step edges.

At other sites of the moir\'{e} pattern, for instance the $top$-$hcp$ site, the incorporation of metastable clusters could be less difficult (no stacking fault is observed in general at these locations). Whether at such sites the metastable cluster (forming a stacking fault) is dissolved before the graphene patch can cover this site, or alternatively, loops of non-hexagonal rings allow to stitch the defect and are subsequently healed, remains unclear at this stage.

Using a lower C adatom concentration during growth or higher temperatures reduces the density of vacant hills, from several 10$^4$~$\mu$m$^{-2}$ to few 10$^3$~$\mu$m$^{-2}$ \cite{natterer2012}. Note that on Rh(111), Ru(0001), and Re(0001), too high temperatures that may allow to fully eliminate the vacant hills become problematic due to the tendency of carbon to dissolve in the bulk \cite{mccarty2009} or to form a surface carbide \cite{dong2012,miniussi2014,qi2017b}. Overall, the vacant hills seem to be traces of the kinetic limitation towards the disappearance of metastable carbon clusters.

\section*{Electronic properties of the vacant hills}

\begin{figure*}[!hbt]
\centering
\includegraphics[width=82.5mm]{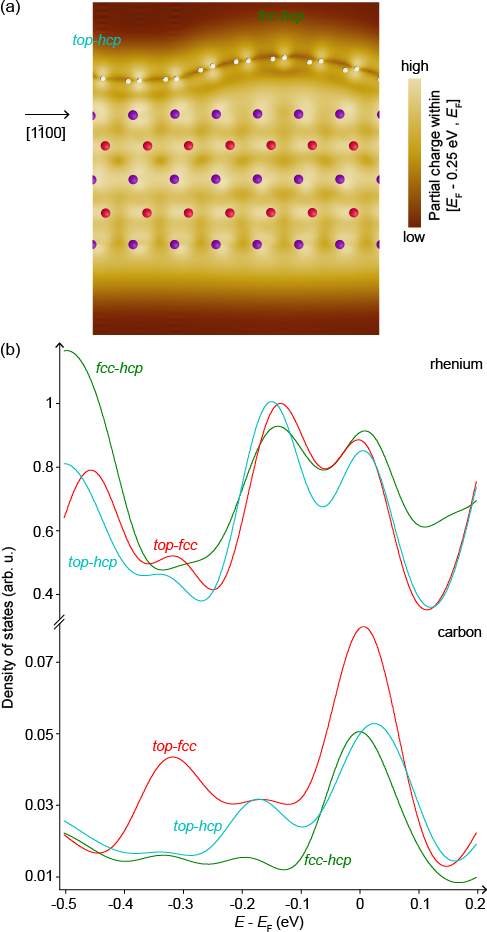}
\caption{(a) Cross-section of the partial charge integrated between the Fermi level ($E_\mathrm{F}$) and $E_\mathrm{F}-0.25$~eV, along the $[1\bar{1}00]$ direction, deduced from the DFT calculations on defect-free graphene/Re(0001); (b) Electronic density of states (electron energy $E$) of individual Re atoms (top curves) and averaged over a few C atoms (four and six respectively, bottom curves) at the hills ($fcc$-$hcp$) and valleys ($top$-$fcc$, $top$-$hcp$) of the moir\'{e}, from DFT calculations.}
\label{fig6}
\end{figure*}

With the help of low temperature STS, we now address the electronic properties of graphene/Re(0001) at different locations in the moir\'{e} pattern and at the location of vacant hills. We find a strong correlation between the local (apparent) height and the local tunneling conductance ($G_\mathrm{t}$) versus tip-sample bias ($V_\mathrm{t}$): the conductance is strongly enhanced 100~meV below the Fermi level, forming a marked peak at the location of the hills ($fcc$-$hcp$) (figure~\ref{fig5}(a)). This peak is absent at the valleys of the moir\'{e} ($top$-$hcp$ and $top$-$fcc$), while at these sites a dip is observed at the Fermi level. The correlation is visualised on figure~\ref{fig5}(b), where the apparent topography is overlaid with the corresponding conductance $G_\mathrm{t}(V_\mathrm{t}=-0.1~\mathrm{eV})$ map.

The conductance spectra at the location of the vacant hill are very similar to those measured at the valleys ($top$-$hcp$ and $top$-$fcc$) of the moir\'{e}, which themselves are hardly discernable from one another. This finding is consistent with our interpretation of the defect as a stacking fault corresponding to either a $top$-$hcp$ or a $top$-$fcc$ stacking, where the graphene-metal distance is similar to that in the moir\'{e} valleys.

To rationalize these observations we turn to the results of DFT calculations. We cannot address the vacant hill itself, since, as mentioned earlier, the supercell that would be required to obtain reliable results would be too large to be tackled. Instead we discuss the spatial variations, within a defect-free moir\'{e} unit cell, of the local density of states, a quantity that contributes to the experimental tunneling conductance. Graphene strongly perturbs the charge distribution at the Re(0001) surface, as can be noticed when comparing the cross sections of the partial charge integrated below the Fermi level, at the graphene/Re(0001) and Re(0001)/vacuum interfaces (respectively top and bottom of figure~\ref{fig6}(a)). The out-of-the-surface extension of the Re wave-functions is indeed strongly reduced at the valleys of the moir\'{e}, due to the hybridization between graphene and Re(0001) orbitals that is the strongest there.

Due to this spatially-varying hybridization, the density of states of the C and Re atoms are themselves varying. In the case of Re atoms, the variation is essentially related to the position in the moir\'{e}, while in the case of C atoms, it is additionally related to the kind of sub-lattice (whether C is on a $top$ or $hcp$ or $fcc$ site with respect to Re atoms makes a large difference). The density of states computed at the hills and valleys of the moir\'{e} is shown in figure~\ref{fig6}(b) (for C atoms, it is averaged over four to six atoms to take into account the contributions of C atoms from the two sub-lattices). Assigning one or the other extrema in these spectra to the ones observed in the experimental conductance is \textit{a priori} risky, since their energy position depends on the degree of approximation that is used for the simulations.

The Re density of states is globally larger than the one of C atoms, but the larger distance to the STM tip is expected to strongly compensate this difference, at least partly. Two maxima are observed for Re and C, both at the hills ($fcc$-$hcp$ region) and the valleys ($top$-$fcc$ and $top$-$hcp$ regions) around the Fermi level and about [-0.2,-0.1]~eV below it (and keeping in mind that the absolute energy values should be taken with caution).

Concerning Re atoms, the lowest-energy maximum is the most intense of the two, and the intensity difference with the other maximum is more severe in the case of the valleys. Concerning C atoms, the higher-energy maximum is the most intense, and the intensity difference with the other maximum is more severe in the case of the hills. Overall, the density of states including the C and Re contributions may hence exhibit marked maxima at the hills (at the higher-energy value), and on the contrary no marked maxima at the valleys (the C and Re contributions somewhat averaging out). These simple arguments elude (possibly important) corrections introduced by the tunneling matrix elements. Yet they suggest that the marked maxima could correspond to the peak measured in the tunneling conductance at the location of hills (figure~\ref{fig5}(a)), and conversely that no specific feature is observed experimentally at the valleys or the vacant hills.

\section*{Summary and concluding remarks}

Using scanning tunneling microscopy/spectroscopy, parametrized bond-order potential calculations and density functional theory electronic structure calculations, we have addressed one of the prominent local defects encountered in epitaxial graphene grown on several metal surfaces such as Re(0001), Ru(0001), and possibly Rh(111) as well. The nature of this defect, which has been observed ever since 2007, had not been discussed in details until now. The defect appears as a vacant hill in the nanorippled pattern that is associated with the moir\'{e} pattern of graphene. This locally, lower-than-expected height and the corresponding lower tunneling conductance both point to a stacking fault, a kind of C-metal stacking that is otherwise encountered at the valleys of the moir\'{e} pattern. Microscopy shows that this vacant hill is incorporated within the surrounding graphene lattice. We discussed the possible structures of the stacking fault, either in the metal surface or in graphene itself, in the latter case in relation with closed loops of non-hexagonal carbon rings surrounding the stacking fault. Our bond-order potential calculations predict a depression in graphene for both kinds of stacking faults, consistently with our observations. We finally proposed that the vacant hills are related to the formation of metastable, kinetically blocked polycyclic clusters forming alongside the growth of graphene.

Whether, as suggested by Lu \textit{et al.} \cite{lu2012}, the vacant hill defect are pathways for the penetration of \textit{e.g.} O atoms and intercalation underneath graphene is an interesting question. Amongst the possible structures of the vacant hill, some include heptagonal and octagonal carbon rings. Through these large rings, small atoms are predicted to penetrate more easily than through the regular hexagons of defect-free graphene \cite{leenaerts2008}. Bond-order potential calculations, that will need to account for the O-C and O-Ru interactions (the construction of the corresponding parametrized potential will require significant efforts well-beyond the scope of the present work), seem a valuable approach to address this question.

In the prospect of producing high-quality graphene, there is an obvious motivation to eliminate the vacant hill defects as much as possible. Ru(0001), Re(0001), and Rh(111) substrates, with which these defects are formed, actually stand out of the broad library of possible graphene substrates because they naturally select a single orientation of graphene, virtually free of other kinds of defects like twin boundaries. Graphene detached from Ru(0001) has already proven high electric transport performance \cite{koren2013}, and we expect that reducing or even eliminating the vacant hills would further improve these performances. Considering metastable polycyclic carbon clusters as a plausible source for the vacant hills, avoiding their formation by \textit{e.g.} the use of \textit{ad hoc} molecular precursors seems a promising route.

With the predicted electronic properties of graphene anti-dot lattices in mind \cite{pedersen2008}, we may change our perspective on the vacant hill defects and now also consider them as exciting objects in themselves. Noting that the moir\'{e} pattern also naturally selects the location of the stacking faults (vacant hills), controlling the number of vacant hills and their 2D organization might allow to engineer unique electronic states with no equivalent in pristine graphene.

\section*{Acknowledgments}
This work was supported by the R\'{e}gion Rh\^{o}ne Alpes (ARC6 program) and the Labex LANEF. We thank A. Kimouche for STM measurements.

\section*{References}

\begin{thebibliography}{10}
\bibitem{noteongrid}
To build the network of lines constituting the grid, we adjust the orientation
  of each individual line so they pass through the atomic positions. This
  triangulation procedure improves the reliability on the determination of the
  atomic positions, and can only be applied in regions where the carbon lattice
  is close-to-ideal.
\end{thebibliography}

\end{document}